\documentclass{aa}

\usepackage{graphicx}
\usepackage{txfonts}

\bibpunct{(}{)}{;}{a}{}{,}

\usepackage{amsmath,amssymb}
\usepackage{subfig}
\usepackage{enumitem}

\newcommand{\be}{\begin{equation}}
\newcommand{\ee}{\end{equation}}

\newcommand{\dd}{\,\mathrm{d}}

\newcommand{\vek}[1]{\mathbf{#1}}

\DeclareMathOperator{\e}{e}

\newcommand{\dalembert}{\Box}

\renewcommand\sun{\odot}

\newcommand*\intd[2][]{
  \mathrm{d}
  \ifx\relax#1\relax\else
  \rule{-0.05em}{1.9ex}^{#1}\!
  \fi
  #2\,
}

\newcommand{\modi}[1]{#1}
\newcommand{\mathmodi}[1]{#1}

\begin{document}

\title{Gravitational redshift profiles in the $f(R)$ and symmetron models}

\institute{Institute of Theoretical Astrophysics, University of Oslo, Postboks 1029, 0315 Oslo\\
  \email{maxbg@astro.uio.no}
}
   \author{Max B. Gronke
     \and Claudio Llinares
     \and David F. Mota}

\date{Version \today}

  \abstract
   {}
   {We investigate the gravitational redshift in clusters of galaxies in the symmetron and Hu-Sawicky $f(R)$ models. The characteristic feature of both models is the screening mechanism that hides the fifth force in dense environments recovering general relativity.}
   {We use $N$-body simulations that were run with the code \texttt{Isis}, which includes scalar fields, to analyse the deviation of observables in modified gravity models with respect to $\Lambda$CDM.}
   {
We find that the presence of the screening makes the deviation highly dependent on the halo mass. For instance, the $f(R)$ parameters $|f_{R0}|=10^{-5}$, $n=1$ cause an enhancement of the gravitational signal by up to $50\%$ for haloes with masses between  $10^{13}\,M_\sun h^{-1}$ and $10^{14}\,M_\sun h^{-1}$.  The characteristic mass range where the fifth force is most active varies with the model parameters.  The usual assumption is that the presence of a fifth force leads to a deeper potential well and thus a stronger gravitational redshift.  However, we find that in cases in which only the central regions of the haloes are screened, there could also be a weaker gravitational redshift.}
   {}


   \maketitle

\section{Introduction}
Although there is overwhelming evidence that dark energy exists, its
nature is still a mystery. One  possible explanation is that  dark
energy is some type of still unknown field. The other is that general
relativity is not fully correct and therefore requires some
modification which can account for the phenomena attributed to dark
energy.
Pursuing the  latter approach, often titled the modification of gravity, intense efforts have been made to establish new theories (see e.g. \citet{Clifton2012} for a theoretical overview). The main problems concerning the development of new gravitational theories are theoretical (e.g. the existence of ghosts) as well as observational, meaning the constraints from local experiments are very tight. One explanation for the possible variations that have not been discovered are so-called screening mechanisms, where the local constraints are fulfilled by hiding the additional force in dense environments such as the solar system.
 
In this work, two different families of scalar-tensor theories with screening are discussed: the symmetron model \citep{Hinterbichler} and an $f(R)$-gravity model that uses chameleon screening \citep{Hu2007}. The first model works through the dependency of the vacuum expectation value of the scalar field on the local matter density, whereas in the latter, the mass of the scalar field is linked to the matter density. Both mechanisms lead to the same outcome. In low-density regions, a fifth force is mediated, and in high-density regions this force is suppressed recovering general relativity. This behaviour requires astrophysical tests in order to examine the existence of such scalar fields. 

Due to the non-linearity of the screening, $N$-body simulations are a preferable tool for observational predictions. For both of these models, this has been done before with the main interest in large-scale imprints, which consist mostly in deviating the matter power spectra and halo mass functions. See \citet{Brax2012}, \citet{Davis2011}, \citet{Winther2011} and, \citet{Llinares2013a} for the symmetron case and \citet{Li2012}, \citet{Brax2013a}, \citet{Li2011} and , \citet{Zhao2011} for the chameleon.  Simulations are also able to predict individual halo properties, such as density or velocity dispersion profiles, which are well established observables.  These observables can be analysed using gravitational lensing \citep{Oguri2012,Mckay2013,Okabe2013}.
Possible constraints for the $f(R)$ model have been established by \citet{Lombriser2012} and \citet{Schmidt2010}.

Another possible test of gravity on scales below $1\,$Mpc is the use of gravitational redshift, since the wavelength shift of light is directly proportional to the depth of the potential well of the clusters of galaxies. This has been observed for the SDSS survey data by \citet{Wojtak2011} who compared the data points to an analytical prediction for general relativity based on the NFW profile \citep{Navarro1995}, an analytical TeVeS prediction\footnote{The analytical TeVeS profile used was later shown to be based on inappropriate assumptions \citep{Bekenstein2012}.}, and a semi-analytical profile from \citet{Schmidt2010}. For the last, the NFW prediction has been boosted by the factor $4/3$, which is the maximum enhancement of gravitational strength according for $f(R)$ gravity.  

The first detection of gravitational redshift in clusters was done by \citet{Wojtak2011} using $7800$ clusters from the SDSS data set. A few corrections to their predictions were made later on \citep{Zhao2012, Kaiser2013} and pointed out that owing to the relative motion of the galaxies, an additional component exists because of time dilation. This term is called the transverse Doppler effect, and it has opposite sign as $v_g$. This and other effects, e.g., a changed redshift as a result of relativistic beaming, have been analysed by \citet{Kaiser2013}, all of them should be considered when doing a proper analysis of observational data. Because the scope of this work is, however, the deviation of $v_g$ in screened modified gravity models with respect to $\Lambda$CDM, these additional effects have not been studied and have not been included in this work. 

In this work, we used data coming from a set of eight N-body simulations that were run with the \texttt{Isis} code covering a variety of model parameters and presented in \citet{Llinares2013}. The box size and resolution were chosen such that it is possible to analyse properties of haloes that correspond to groups and clusters of galaxies.  We study the gravitational redshift profiles in the objects found in the simulations in the three gravitational models and show that such comparisons and predictions have to be made with extreme caution when dealing with screening models.  During the analysis, special emphasis was put on determining the virialisation state of the haloes by taking the energy of the scalar field into account.

The outline of the paper is as follows. In Sec.~\ref{sec:models} we review the concept of scalar-tensor theories with screening and with particular focus on the symmetron and the Hu-Sawicky $f(R)$-model. In Sec.~\ref{sec:data_set} we describe the simulations and the methods employed for extracting information from the data sets. In particular, the halo selection and virialisation are discussed. In Sects.~\ref{sec:results} and \ref{sec:discussion} we present and discuss the results. Finally, we conclude in Sec.~\ref{sec:conclusions}.

\section{Models}
\label{sec:models}

This section gives a general description of scalar-tensor theories with screenings mechanisms.  Furthermore, the particular models that are the subjects of this paper are briefly reviewed.  The implementation of the equations in the N-body code that was used to run the simulations is described in detail in \citet{Llinares2013}.

\subsection{Scalar-tensor theories with screening}
\label{sec:scalar-tensor-w-screening}

The action of a scalar field $\phi$ in a scalar-tensor theory is given by
\be
S = \int\dd^4x \sqrt{-g}\left(\frac{M_{\mathrm{Pl}}^2}{2}R-\frac{1}{2}\nabla_\mu\phi\nabla^\mu\phi-V(\phi)\right) + \mathcal{S}_m(\psi^{(i)},\tilde g_{\mu\nu})
\ee
where the matter fields $\psi^{(i)}$ couple to the Jordan frame metric given by
\begin{equation}
  \tilde g_{\mu\nu}\equiv A^2(\phi)g_{\mu\nu}.
\end{equation}
This leads to a equation of motion for the scalar field
\begin{equation}
  \label{eq:scalar_field_eom}
  \dalembert\phi=V'(\phi)-A'(\phi) T
\end{equation}
where $T$ is the trace of the energy-momentum tensor. The right hand-side of this equation can be identified with an effective potential:
\begin{equation}
  V_{\text{eff}}=V(\phi)-(A(\phi)-1)T.
  \label{eq:effective_potential}
\end{equation}
As in a matter-dominated background $T\approx -\rho_m$, the value of the minimum depends on the local matter density. 
The screening nature of a model is now given if $V_{\text{eff}}$ has a minimum close to zero for high-density backgrounds.

To describe the behaviour of the field in the environment of matter, we use the Newtonian gauge metric given by
\begin{equation}
  \dd s^2 = -(1+2\Phi)\dd t^2 + a^2(t)(1-2\Phi)(\dd x^2+\dd y^2+\dd z^2).
\end{equation}
The geodesic equation in this particular coordinates becomes
\begin{equation}
  \vek{\ddot x}+2H\vek{\dot x} +\frac{1}{a^2}\vec\nabla\left(\Phi+\log A\right) = 0
  \label{eq:geodesic_general}
\end{equation}
where the quasi-static limit has been applied (i.e. the time derivatives of the field were assumed to be much smaller than its spatial variation). The last term of the equation can be interpreted as an effective force potential, and the arising fifth force is given by
\begin{equation}
  \vek{\ddot x}_{\mathrm{Fifth}} = -\frac{1}{a^2}\vec\nabla\log A.
  \label{eq:ddot_x_fifth}
\end{equation}

\subsection{Symmetron}

In the symmetron model \citep{Hinterbichler,Hinterbichlera,Olive2008}, the expectation value of the scalar field depends on the local matter density. In high-density regions it is zero, but when the matter density is lower than a given threshold, the symmetry of the potential is broken, and the potential acquires a minimum away from zero, leading to an additional force in these regions.

The requirement for the symmetron is that both $A(\phi)$ and $V(\phi)$ are symmetric under $\phi\rightarrow -\phi$, and therefore a commonly considered coupling is
\begin{equation}
  A(\phi)=1+\frac{\phi^2}{2M^2}+\mathcal{O}\left(\frac{\phi^4}{M^4}\right)
\end{equation}
where $M$ is a mass scale.  The simplest potential can be stated as
\begin{equation}
  V(\phi)=-\frac{1}{2}\mu^2\phi^2+\frac{1}{4}\lambda\phi^4, 
\end{equation}
where $\mu$ is a mass scale and $\lambda$ a dimensionless parameter.  This definitions leaves the effective potential (\ref{eq:effective_potential}) for non-relativistic matter as 
\begin{align}
  V_{\mathrm{eff}}&=\frac{1}{2}\left(\frac{\rho}{M^2}-\mu^2\right)\phi^2+\frac{1}{4}\phi^4,
\end{align}
which has a minimum at $\phi=0$ for densities $\rho>\rho{ssb}\equiv \mu^2 M^2$ and two minima if $\rho<\rho_{ssb}$. Particularly, the two minima are $\phi_0=\pm\mu/\sqrt{\lambda}$ in vacuum.

The geodesic equation for this model turns into
\begin{equation}
  \vek{\ddot x} + 2 H\vek{\dot x} + \frac{1}{a^2}\nabla\Phi + \frac{1}{(M a)^2}\phi\nabla\phi = 0,
  \label{eq:geodesic_symmetron}
\end{equation}
hence, $\vek{\ddot x}_{\text{Fifth}}\propto \phi\nabla\phi$. Consequently, in a high-density region, the fifth force is screened as required.

Similar to \citet{Winther2011}, instead of the original parameters $(\mu, M, \lambda)$, more physical parameters were introduced that are all linked to the properties of the scalar field in vacuum. Firstly, the range of the field in vacuum
\begin{equation}
  L = \frac{1}{\sqrt{2}\mu};
\end{equation}
secondly, the expansion factor for which the symmetry is broken in the background level
\begin{equation}
  a_{ssb}^3 = \frac{\Omega_{m0}\rho_{c0}}{\mu^2 M^2};
\end{equation}
and finally a dimensionless coupling constant
\begin{equation}
  \beta = \frac{\mu M_{Pl}}{M^2 \sqrt{\lambda}}.
\end{equation}
In addition, the scalar field itself is normalised with its vacuum expectation value:
\begin{equation}
 \chi\equiv \phi/\phi_0 = \frac{\phi a_{ssb}^3}{6 H_0^2 M_{Pl} L^2 \beta \Omega_{m0}}.  
\end{equation}
With these redefinitions, the geodesic equation can be rewritten to
\begin{equation}
  \vek{\ddot x} + 2 H \vek{\dot x} + \frac{1}{a^2}\nabla\Phi + \frac{6 H_0^2 L^2 \beta^2}{a^2 a_{ssb}^3}\chi\nabla\chi = 0.
  \label{eq:geodesic_isis_symmetron2}
\end{equation}
The equation of motion of the scalar field is in this case
\begin{equation}
  \nabla^2\chi = \frac{a^2}{2 L^2}\left[\left(\frac{a_{ssb}}{a}\right)^3\eta\chi - \chi + \chi^3\right]
\end{equation}
where $\eta\equiv \rho_m a^3/(\Omega_{m0}\rho_{c0})$ is the matter density in terms of the background density.

\subsection{Hu-Sawicky $f(R)$-model}

In the $f(R)$ model by \citet{Hu2007}, the Jordan frame action is given by
\begin{align}
  S&=\int\dd^4x\bigg\{\sqrt{-\tilde g}\left(\tilde R+f(\tilde R)\right)+\mathcal{L}_m(\psi^{(i)},\tilde g_{\mu\nu})\bigg\}, 
  \label{eq:fofr_hu_sawicky}
\end{align}
where
\begin{equation}
f(\tilde R) = -m^2\frac{c_1(\tilde R/m^2)^n}{1+c_2(\tilde R/m^2)^n}.  
\end{equation}
Here, $c_1$, $c_2$, $m$, and $n$ are positive constants and $m$ is specified as $m^2 = H_0^2/\Omega_{m0}$. Demanding $\Lambda$CDM background evolution ($c_1/c_2 m^2=2\Lambda$) and taking the high curvature limit, which is an expansion around $m^2/\tilde R\rightarrow 0$, and introducing the current background curvature $f_{R0}\equiv \dd f/\dd \tilde R\big|_{a=a_0}$, Eq.~(\ref{eq:fofr_hu_sawicky}) can be written as
\begin{equation}
  f(R) = - 16\pi G \rho_{\Lambda} - \frac{f_{R0}}{n}\frac{\tilde R_0^{n+1}}{\tilde R^n}.
  \label{eq:fofr_eq3}
\end{equation}
The $f(R)$ models can be
treated as scalar-tensor theories, via a conformal transformation:
\begin{equation}
  \tilde g_{\mu\nu}=\e^{2\beta\phi/M_{Pl}}g_{\mu\nu}\quad\text{with }\beta=\frac{1}{\sqrt{6}}.
\end{equation}
Under this transformation to the Einstein frame metric $g_{\mu\nu}$, a scalar field appears with
\begin{equation}
  f_R = \e^{-2\beta\phi/M_{Pl}}-1\approx -\frac{2 \beta \phi}{M_{Pl}},
  \label{eq:dfdr_scalar}
\end{equation}
and its potential is given by
\begin{equation}
  V(\phi)=M_{Pl}^2\frac{f_R \tilde R-f}{2(1+f_R)^2}.
  \label{eq:fofr_potential}
\end{equation}
The effective linearised potential reads as
\begin{align}
  V_{\text{eff}} &= V(\phi) + \frac{\beta\phi}{M_{Pl}}\rho.
\end{align}

By inserting the above definitions in Eq.~\eqref{eq:geodesic_general}, the geodesic equation for this model can be written as
\begin{equation}
  \vek{\ddot x}+2H\vek{\dot x} +\frac{1}{a^2}\vec\nabla\left(\Phi-\frac{f_R}{2}\right) = 0.
  \label{eq:geodesic_fofr_code}
\end{equation}
Hence, the additional force in this case is $\vek{\ddot x}_{Fifth}\propto \nabla f_R\propto \nabla \phi$. Since the minimum of $V_{\text{eff}}$ approaches zero with a growing $\rho$, the fifth force is screened in high-density regions as seen before.

The parameters of the Hu-Sawicky model discussed before, namely $n$ and $f_{R0}$, were also used in the simulation. Basically, the effective potential and the conformal transformation described can be used directly to solve the modified geodesics and the scalar field evolution numerically as done for the symmetron.  Sin the scalar field has a fixed sign in this case, it is customary to stabilise the numerical solution by forcing its sign to be unique \citep{Oyaizu2008}.  This can be made by including a new change of variables of the following form:
\begin{equation}
  \e^{u}\equiv -f_R a^2.
\end{equation}
This leads to the equation of motion for the $f_R$-field
\begin{align}
&\nabla\cdot\left(\exp(u)\nabla u\right) =\Omega_m a H_0^2(\tilde{\rho}-1)\nonumber\\
&- \Omega_m a H_0^2\left(1 + 4\frac{\Omega_\Lambda}{\Omega_m}\right)(|f_{R0}|a^2)^{\frac{1}{n+1}}e^{-\frac{u}{n+1}}\nonumber\\
&+\Omega_m a H_0^2\left(1 + 4a^3\frac{\Omega_\Lambda}{\Omega_m}\right).
\end{align}

\section{Simulations and analysis}
\label{sec:data_set}

\subsection{The simulations}

The simulations that we used for the analysis were run with the code \texttt{Isis} \citep{Llinares2013}, which is a modification of \texttt{RAMSES} \citep{Teyssier2002} which includes the development of the scalar fields.  All simulation runs contained $512^3$ dark-matter-only particles with a mass of $9.26138\times 10^9\,M_{\sun}/h$.  The background cosmology is defined as $(\Omega_{m0}, \Omega_{\Lambda 0}, H_0)=(0.267,\,0.733,\,71.9)$. The simulation box has a side length at redshift zero of $256\,h^{-1}$Mpc and the boundary conditions are periodic.  The data sets are the snapshots taken at $z=0$.  All simulations were run with the same initial conditions, which were generated  using Zeldovich approximation with standard gravity with the package Cosmics \citep[][]{1995astro.ph..6070B}.  In doing this, it was assumed that both extended models give fully screened fields before the initial redshift of the simulation.  We refer the reader to  \citet{Llinares2013} for further details on the simulations. 

\begin{table}
  \caption{Model parameters of the different simulation runs.}
  \centering
\subfloat{
  \begin{tabular}{@{}lrc@{}}
    \hline \hline
    Name & $|f_{R0}|$ & $n$ \\
    \hline
    fofr4 & $10^{-4}$ & $1$ \\
    fofr5 & $10^{-5}$ & $1$ \\
    fofr6 & $10^{-6}$ & $1$ \\
    \hline
  \end{tabular}
}
\qquad
\subfloat{
  \begin{tabular}{@{}lrrc@{}}
    \hline \hline
    Name & $z_{ssb}$ & $\beta$ & $L$ \\
         &          &         & (Mp\modi{c} $h^{-1}$) \\ 
    \hline
    symm\_A & $1$ & $1$ & $1$ \\
    symm\_B & $2$ & $1$ & $1$ \\
    symm\_C & $1$  & $2$ & $1$ \\
    symm\_D & $3$ & $1$ & $1$ \\
    \hline
  \end{tabular}
}
\label{tab:run_params}
\end{table}

The parameters for the $f(R)$ and the symmetron models are summarised in Table~\ref{tab:run_params}.  In the case of the $f(R)$-gravity, the parameter $n$ was fixed to one, while $f_{R0}$ took values from $10^{-6}$, which resulted in hardly any deviation from $\Lambda$CDM, to $10^{-4}$, which is on the border of violating cluster abundance constraints \citep{Ferraro2011}.  For the symmetron model mainly the time of the symmetry breaking was varied while leaving the others constant except for one of the model that also has an increased coupling constant.

\subsection{Halo selection}
\label{sec:halo_selection}
The halo identification was made using the halo finder \texttt{Rockstar} \citep{Behroozi2013}, a publicly available 6D FOF code. The boundary of the haloes was chosen to be $R_{200c}$ (i.e. where the density falls below $v=200$ times the critical density today). A proper definition of halo properties in screened modified gravity would require $v$ to vary with the halo mass and the local environment, as spherical collapse analysis shows \citep{Brax2010,Li2012a}. We found, however, that our results are stable against the change in $v$ for the modified gravity models. For further analysis, we refer to this halo boundary as virial radius (i.e. $R_v\equiv R_{200c}$). Since we are dealing with gravitational effects, for which the total mass is the important quantity, we include the subhaloes of a given host as part of the main halo throughout the analysis.  We defined the centre of the objects as the position of the particle that corresponds to the minimum gravitational potential.  This choice is aimed at getting the difference in gravitational potential between the brightest cluster galaxy (BCG) and the rest of the cluster, thereby imitating an observational standpoint.  This can, however, be a problem because the minimum of the gravitational potential is not necessarily the densest central region. To verify the findings, the results were also reproduced by taking the halo definition \modi{of the biggest subhalo} from the \texttt{Rockstar} halo finder, and only minor deviations inside the virial radius were found.

The $\Lambda$CDM run contains around $73,000$ dark matter haloes with more than $100$ particles and circa $9200$ with more than $1000$ particles.  The rest of the simulations have similar mass functions with small differences that come from the existence of the scalar field\footnote{\modi{The number count in the considered mass bins was increased by $\sim 15$\%. The greatest deviation was found in the \textit{symm\_D} model with $\sim 35$\% throughout the mass bins.}}. An in-depth interpretation of the change in the halo mass functions in the symmetron and chameleon model was done by \citet{Brax2012,Brax2013a} and is not the subject of this work.

\subsection{Virialisation}
\begin{figure*}
  \includegraphics[width=17cm]{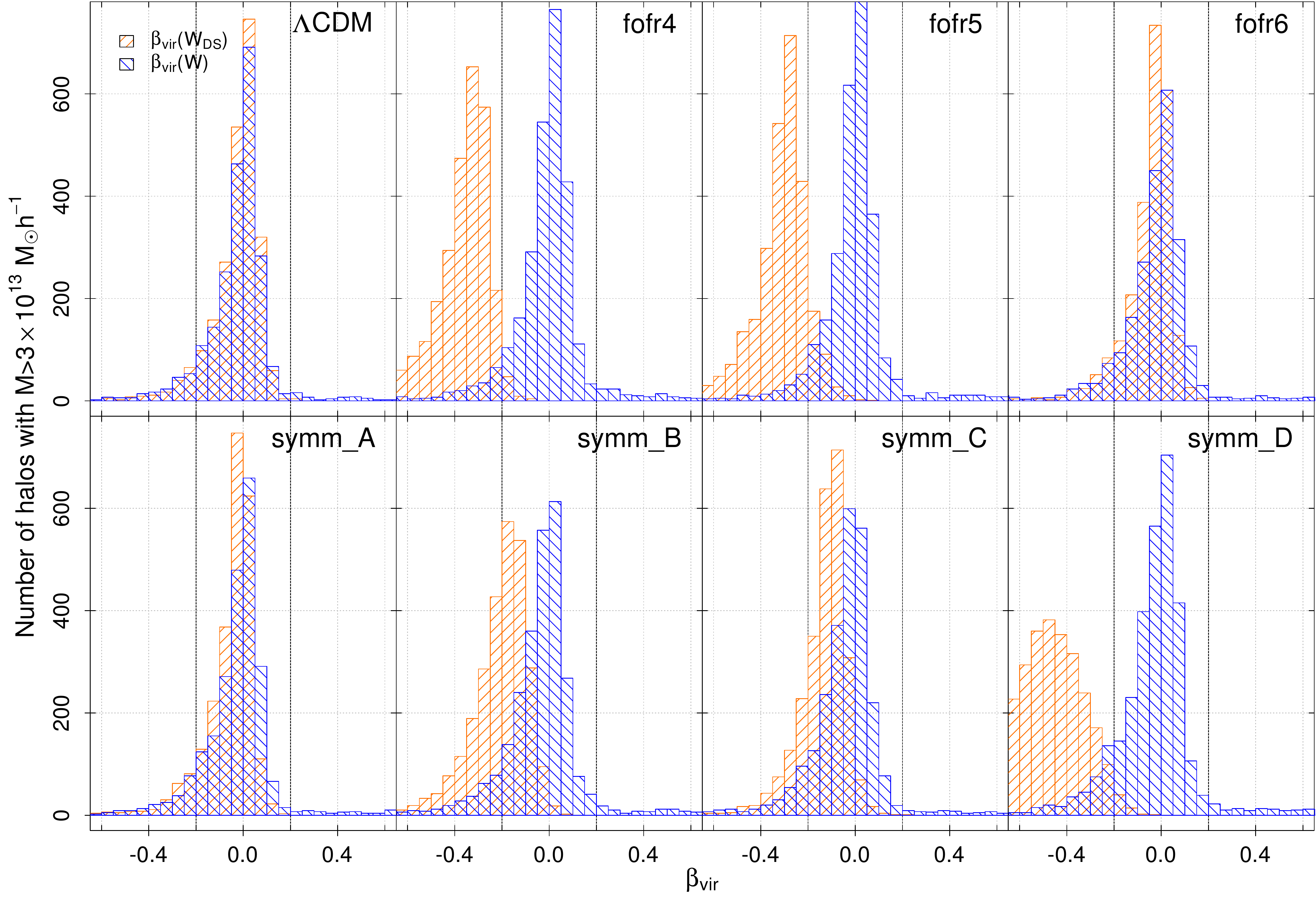}
  \caption{Distribution of the virialisation parameter $\beta_{vir}$ for haloes with mass $M>3\times 10^{13}M_\sun h^{-1}$. $\beta_{vir}$ using the potential energy obtained through direct summation $W_{DS}$, which was defined in Eq.~\eqref{eq:W_DS}, is shown in orange with upward hatching, and $\beta_{vir}$ using $W$, defined in Eq.~\eqref{eq:W_part}, is shown in blue with downward hatching. \modi{The vertical black lines visualise our criterion for virialisation, i.e. $|\beta_{vir}|<0.2$.}}
  \label{fig:hist_vir_compare}
\end{figure*}
To study halo properties, it is crucial not to mix haloes that are dynamically relaxed and those that have not reached such an equilibrium state yet because the two groups have a different distribution of halo properties \citep{Shaw2006}. Using the whole cluster sample and not separating these two groups may therefore lead to a skewed mean and higher noise. Especially when comparing the effect of different gravitational theories below Mpc scale -- as done in this work -- combining the two groups may lead to a false interpretation because it is not clear if the change happens on an actual halo property or if only the number of unrelaxed objects has changed.

By definition the energy portions of a virialised object fulfil the virial theorem:
\begin{equation}
  2T + W - E_S = 0
\end{equation}
with the potential energy $W$, the kinetic energy $T$, and the surface pressure term $E_s$. For a perfect fluid with density $\rho$ and pressure $p$, these quantities are given by
\begin{equation}
  W = \int_V\intd[3]{\vek{x}}\rho\vek{x}\cdot\vek{\ddot x} \label{eq:W_fluid}\;,
\end{equation}
\begin{equation}
E_S = \int\limits_S\dd\vek{S}\cdot(\vek{x}p) \label{eq:E_S_fluid}
\end{equation}
and
\begin{equation}
T = \frac{3}{2}\int\limits_V\intd[3]{\vek{x}}p. \label{eq:T_fluid}
\end{equation}
Since these relations are derived from the collisionless Boltzmann equation \citep[e.g.][]{Chandrasekhar1981}, they are true in any modified gravity model, hence also in the models we consider. Any modification in gravitational force enters through the total acceleration $\vek{\ddot x}$, which is a sum of the Newtonian part and an acceleration induced by a fifth force.

The previous definitions must be discretised if one wants to apply them to N-body simulations.   The kinetic energy term then yields
\begin{equation}
 T = \frac{1}{2}\sum\limits_i m_i\vek{v}_i^2
\end{equation}
where $\vek{v}_{i}$ is the relative velocity of particle $i$ with respect to the halo velocity.

The potential energy was obtained using the acceleration of the particles:
  \begin{align}
  W &= \sum\limits_{i}\frac{m_i}{2}(\vek{x_i}-\vek{x_{\mathcal H}})\cdot\vek{\ddot x_i}, 
  \label{eq:W_part}
\end{align}
where the centre of the halo $\vek{x_{\mathcal H}}$ was defined as the minimum of the gravitational potential. Since the acceleration was calculated using the Newtonian and the fifth force, the definition ensures that the energy of the scalar field is not neglected. Both forces were obtained directly through the $N$-body code via an inverse cloud-in-cell (inverse CIC) smoothing, in the same way as done while running the simulations.

The $E_S$ term was calculated following \citet{Shaw2006}, who use an approximation that is motivated by the ideal gas law and reads as
\begin{equation}
E_s\approx 4\pi R_{\mathmodi{90}}^3 p_S,
\end{equation}
where
\begin{equation}
  p_S = \frac{1}{3}\frac{\sum_i m_i\vek{v}_i^2}{4/3\pi(R_{\mathmodi{100}}^3-R_{\mathmodi{80}}^3)}.
\end{equation}
Here, the sum in the nominator was carried out using the outermost $20\%$ of the particles within $R_{vir}$, and the notation $R_{x}$ is used to describe the $x$-percent quantile of this particle distribution.

Measuring the level of relaxation can be done using the virialisation parameter, which is defined as
\begin{equation}
  \beta_{vir}\equiv \frac{2 T - E_s}{W} + 1.
\end{equation}
We then defined a halo as being sufficiently relaxed if $|\beta_{vir}|<0.2$.  This removes around half of all the haloes from each data sample. Mainly smaller haloes are removed. For the $\Lambda$CDM data, the three halo mass bins that will be adopted later on (i.e. $\log(M h/M_\sun) \in ((13,\,13.5),~(13.5,\,14),~(14,\,14.5))$), loose $(19,~13,~18)$ percent of the haloes, which leaves $(5060,~1544,~314)$ haloes. The three haloes in the $\Lambda$CDM data set with masses higher than $10^{14.5}M_\sun\,h^{-1}$ are also unvirialised according to our definition. \modi{These rather large quantities of removed haloes are mainly due to our previously introduced halo definition after which subhaloes are merged into the host halo, so that the full phase-space analysis of \texttt{Rockstar} is partially lost.}

Figure~\ref{fig:hist_vir_compare} shows the distribution of the virialisation parameter for haloes with masses $M>3\times 10^{13}M_\sun h^{-1}$. To illustrate that the energy of the scalar field must not be neglected, $\beta_{vir}$ has been calculated not only with a potential energy as stated in Eq.~\eqref{eq:W_part} but also with a potential energy obtained through direct summation using \textit{only} standard gravity:
\begin{equation}
  W_{DS} = -G\sum\limits_{ij}\frac{m_i m_j}{|\vek{x_i}-\vek{x_j}|}. 
  \label{eq:W_DS}
\end{equation}
For $\Lambda$CDM, the \textit{fofr6} or the \textit{symm\_A} model, the deviation is not significant, but in the more extreme models (\textit{fofr4}, \textit{symm\_D}) the direct summation method is clearly not sufficient. For them, the higher kinetic energy due to additional acceleration is not compensated through an equally increased potential energy and the $\beta_{vir}$-distribution is shifted to lower values.

\subsection{Measuring the gravitational redshift from the simulations}
In order to obtain the gravitational redshift, first the gravitational potential at each \modi{of the} particles' positions is needed. \modi{To achieve this, the $N$-body code was modified to interpolate several properties into the particles' positions and output them. Amongst these properties were, for example, the matter density and the gravitational potential.}

The measure of gravitational redshift was then defined as
\begin{equation}
  v_g = \frac{\Delta\Phi}{c} = \frac{\Phi_c - \Phi}{c}.
  \label{eq:v_g}
\end{equation}
Here, $\Phi_c$ is the minimum gravitational potential.  The particle that corresponds to this value was used (as in previous section) to define the centre of the halo.  Therefore, $v_g$ is always negative and can be interpreted as the gravitational blueshift of light seen by an observer located at the centre of the halo.

Instead of defining a single point as centre of the cluster and calculating the gravitational redshift with respect to that point, it is also possible to model the central galaxy as being spread out and take an average over gravitational potentials of most central particles as the reference point. This procedure is used by \citet{Kim2004} and leads to a flattening of the profile. However, since they found out the effect is rather weak for dark-matter-only simulations, and our results are consistent with the NFW predictions we do not adopt the alternative procedure.

\modi{Our approach yields values for $v_g(R_{vir})$ of approximately $-7\,$km/s and $-1.5\,$km/s in the mass bins $10^{13}-10^{13.5}M_\sun/h$ and $10^{14}-10^{14.5}M_\sun/h$, respectively. This is consistent with the findings of \citet{Wojtak2011} and \citet{Kim2004}, keeping in mind their altered halo mass bins.}

\begin{figure*}
  \centering
  \includegraphics[width=17cm]{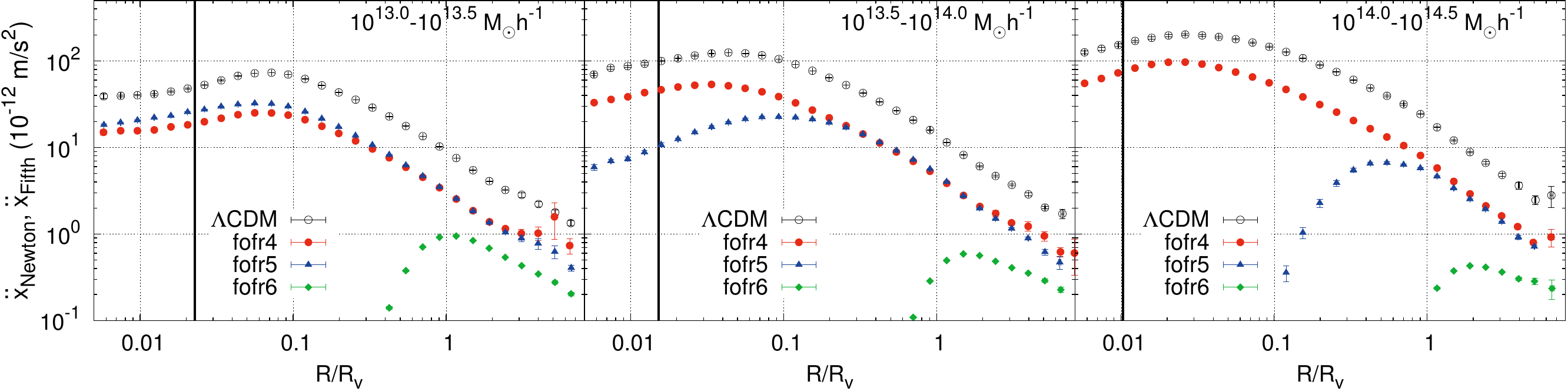}
  \\
  \includegraphics[width=17cm]{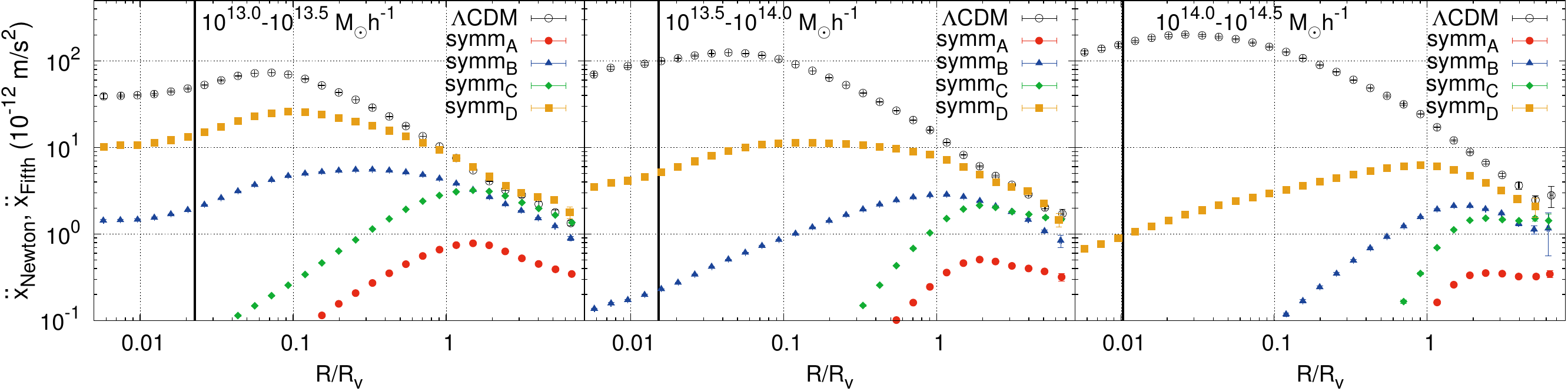}
  \caption{Additional acceleration $|\vek{\ddot x}|_{\mathrm{Fifth}}$ due to the presence of a scalar field in the $f(R)$ (top) and symmetron (bottom) models as a function of radius for three different mass bins. \modi{Also, the pure Newtonian gravitational force $|\vek{\ddot x}|_{\mathrm{Newton}}$ is included from the $\Lambda$CDM data set (black, unfilled circles) as a comparison.} The vertical line approximately corresponds to two grid cells in the finest refinement level.  See text for details.}
  \label{fig:fifth_force_radius}
\end{figure*}

\section{Results}
\label{sec:results}

 The aim of the paper is to study possible differences in the gravitational redshift signal owing to the presence of a scalar field.  However, since null geodesics are invariant under conformal transformations and, thus, identical in the Einstein and Jordan frame, the photons are not affected by the presence of scalar fields. Consequently, the gravitational redshift is not influenced directly by the scalar field, but maps out the change in matter clustering due to the presence of the fifth force.

To understand the change in the gravitational redshift profiles, we also analysed the magnitude of the fifth force.  For each of the quantities under study, we obtained an estimation of the error by dividing the simulation box into eight sub-boxes and using the variance of the relative deviation calculated in each subset to fix the error of the mean.

\begin{figure*}
  \centering
  \includegraphics[width=17cm]{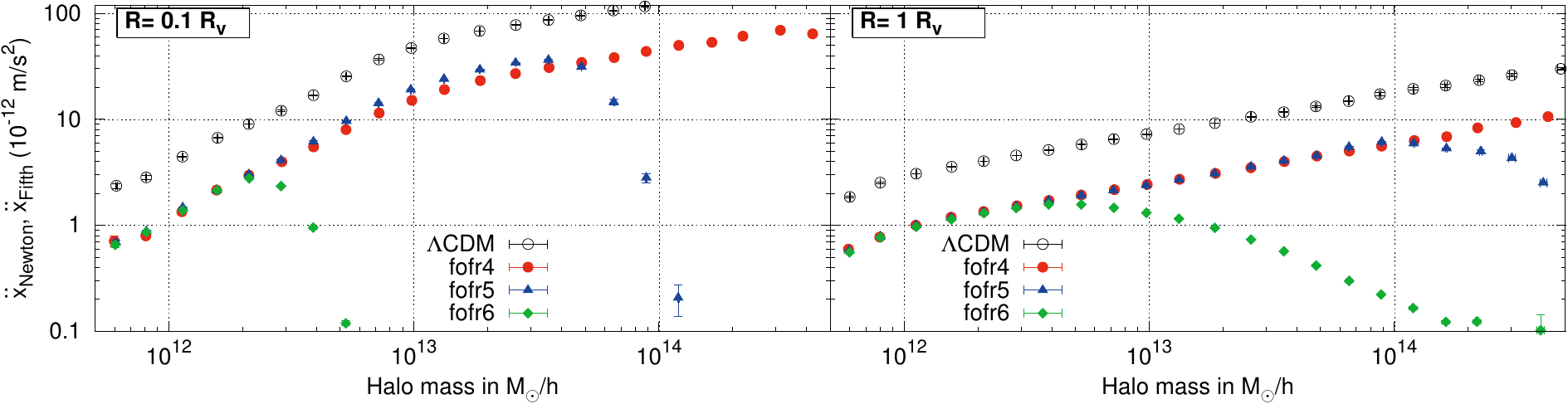}
  \\
  \includegraphics[width=17cm]{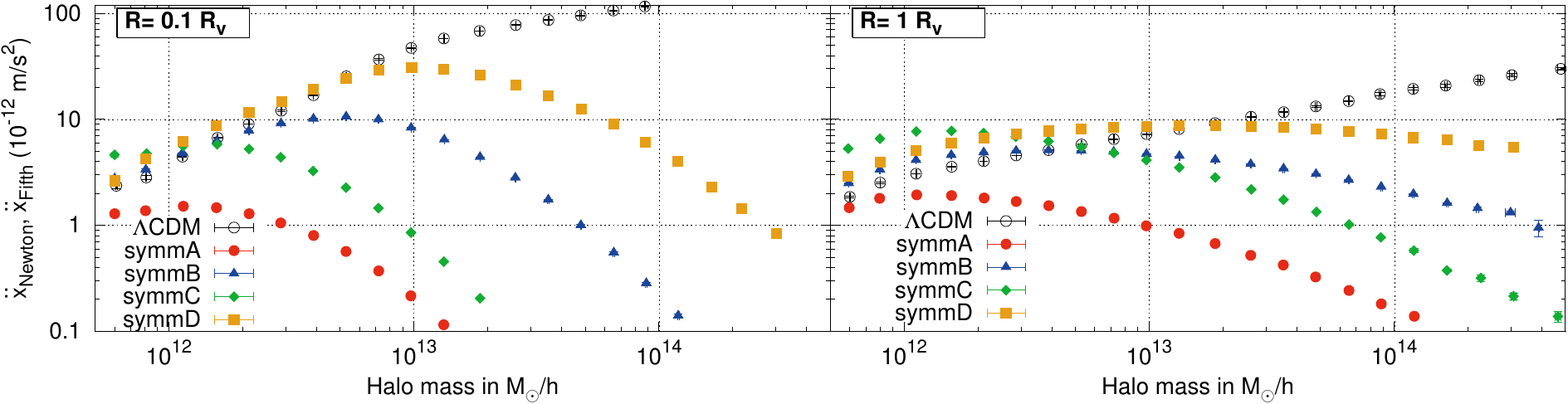}
  \caption{Additional acceleration $|\vek{\ddot x}|_{\mathrm{Fifth}}$ due to the presence of a scalar field in the $f(R)$ (top) and symmetron (bottom) models as a function of halo mass for two different radial bins. \modi{As in Fig.~\ref{fig:fifth_force_radius}, the black unfilled circles represent $|\vek{\ddot x}|_{\mathrm{Newton}}$ as a comparison.}}
  \label{fig:fifth_force_mass}
\end{figure*}

\subsection{Distribution of fifth force in the dark matter haloes}
Figure \ref{fig:fifth_force_radius} shows the absolute value of the acceleration $|\vek{\ddot x}_{\mathrm{Fifth}}|$ defined by Eq.~\ref{eq:ddot_x_fifth} as a function of radius at redshift $z=0$.  The forces and scalar field that are needed to evaluate $|\vek{\ddot x}_{\mathrm{Fifth}}|$ were calculated in the same way as we did when calculating the virialisation state (i.e. using the same smoothing kernel that the $N$-body code \texttt{Isis} used during the simulations to interpolate quantities from the grid into the particle's position).  The adaptive mesh refinement of the code was used during these calculations, and so we reach the same resolution as during the simulations.  The upper and bottom rows correspond to $f(R)$ and symmetron simulations, respectively.  We show results for three different mass bins for each model.  The vertical line to the left of the panels corresponds to the resolution limit that was estimated as twice the size of the grid of the deepest refinement level normalised with the mean virial radius of each mass bin.  

All the curves show a characteristic maximum, whose position is highly dependent on the model parameters and the mass of the haloes.  In the $f(R)$ case, the maximum of the fifth force profile moves towards larger radii when increasing mass  and decreasing the only free parameter $|f_{R0}|$.  This is a direct consequence of the screening that is activated in a larger part of the haloes when decreasing $|f_{R0}|$.  For the symmetron model, an increase of $\beta$ or $z_{SSB}$ leads to a stronger fifth force. A greater $\beta$ value increases the $|\vek{\ddot x_{\text{Fifth}}}|$ values by a constant factor (\textit{symm\_A} versus \textit{symm\_C}) while altering $z_{ssb}$ changes the shape of the fifth force profile in general.

To better understand the mass dependence of the distributions, we show in Fig.~\ref{fig:fifth_force_mass} the absolute value of the acceleration $|\vek{\ddot x}_{\mathrm{Fifth}}|$ as a function of halo mass as measured at two different radii. The information presented in the plots was obtained from two spherical shells of radius $0.1 R_v$ and $R_v$ and a thickness of $0.02 R_v$.  The upper panels of the figure show the results from the $f(R)$ simulations, while the bottom panels correspond to the symmetron data. From this figure, it is clear that different sized haloes are variably affected by the fifth force. The model parameters lead to a characteristic halo mass range beyond which the fifth force is screened.  Below this range, the haloes are not enough dense for the fifth force to be activated.  In the case of the $f(R)$ model, we find that the low mass end of the force distribution is completely insensitive to changes in the only free parameter $f_{R0}$.  On the contrary, we find a strong dependence in the high mass end of the distribution:  the larger $f_{R0}$, the higher the masses that are screened and thus affected by the fifth force.  This produces a shift in the maximum of the distributions towards higher masses when $f_{R0}$ is increased.

The dependence of the symmetron fifth force on the redshift of symmetry breaking $z_{SSB}$ is similar to the dependence of this quantity on $f_{R0}$ in the Hu-Sawicky model:  the higher the symmetry breaking, the higher the masses that are unscreened and affected by the fifth force.  Since the low mass end of the distribution is also rather insensitive to changes in $z_{SSB}$, there is a displacement of the maximum of the distributions towards high masses when increasing $z_{SSB}$.  By comparing simulations \textit{symm\_A} and \textit{symm\_C}, we can test the dependence of the forces when changing the coupling constant $\beta$.  We find a rather insensitive behaviour.  The only changes are in the normalisation and are given by the dependence with $\beta$ in equation \ref{eq:geodesic_isis_symmetron2}.

\begin{figure*}
  \centering
  \includegraphics[width=.95\textwidth]{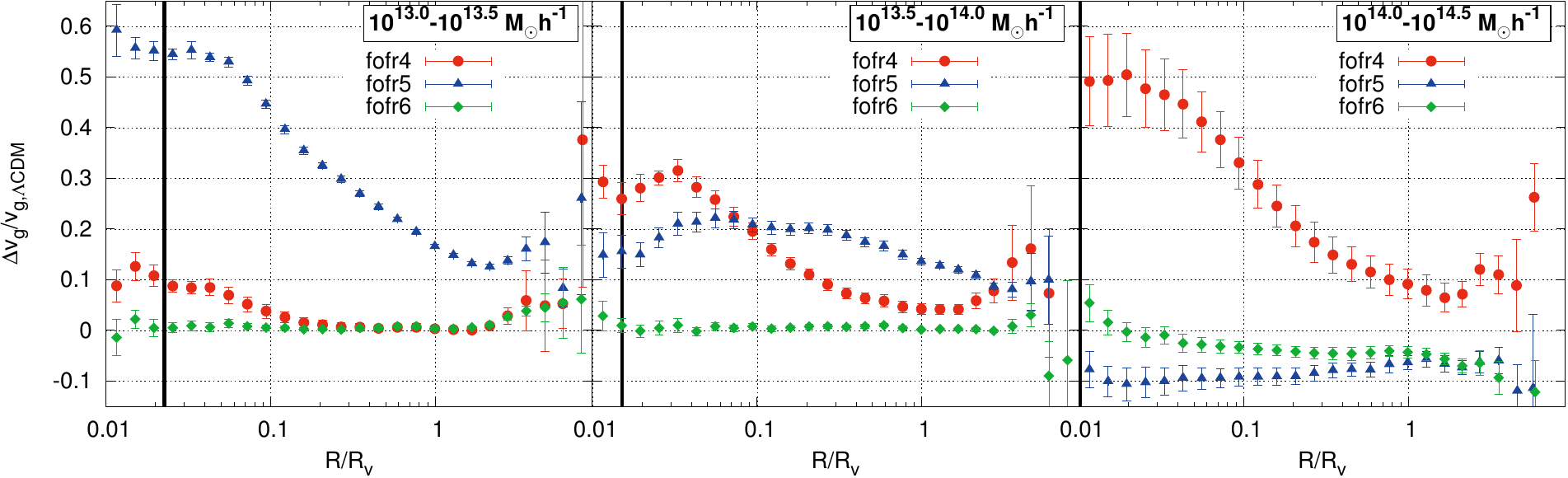}  
  \\
  \includegraphics[width=.95\textwidth]{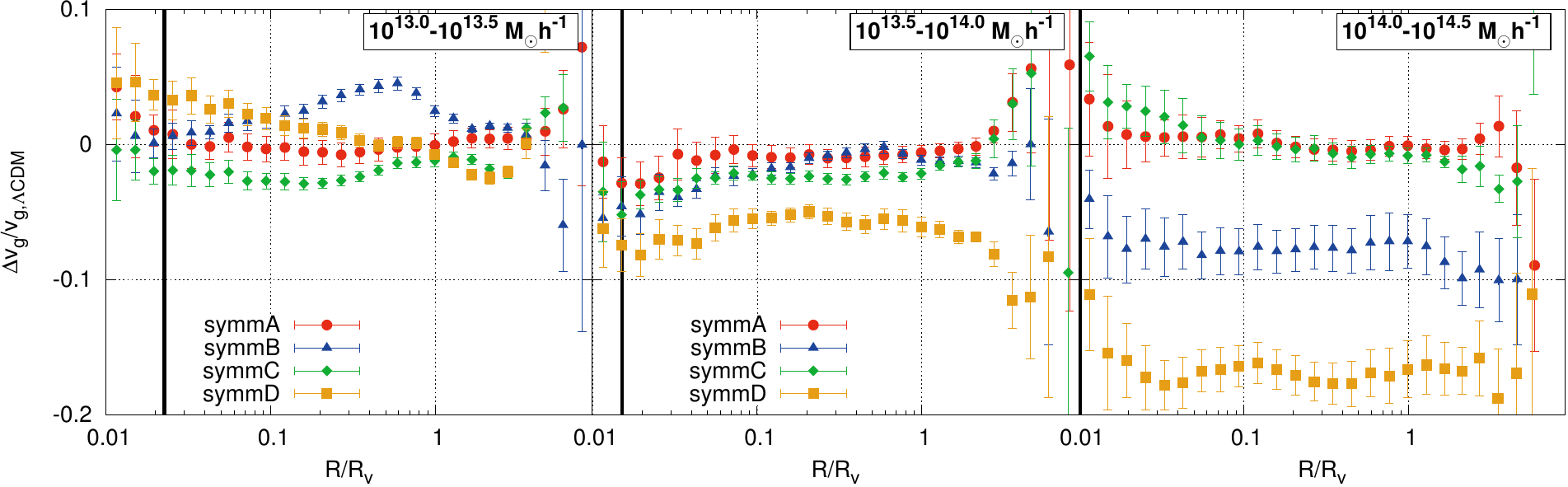}
  \caption{Relative deviation in gravitational redshift between the scalar-tensor models and $\Lambda$CDM as a function of radius. The upper panels correspond to the $f(R)$ model and the lower to the symmetron. The vertical line approximately corresponds to two grid cells in the finest refinement level.}
  \label{fig:grav_red_radius}
\end{figure*}

\begin{figure*}
  \centering
  \includegraphics[width=.95\textwidth]{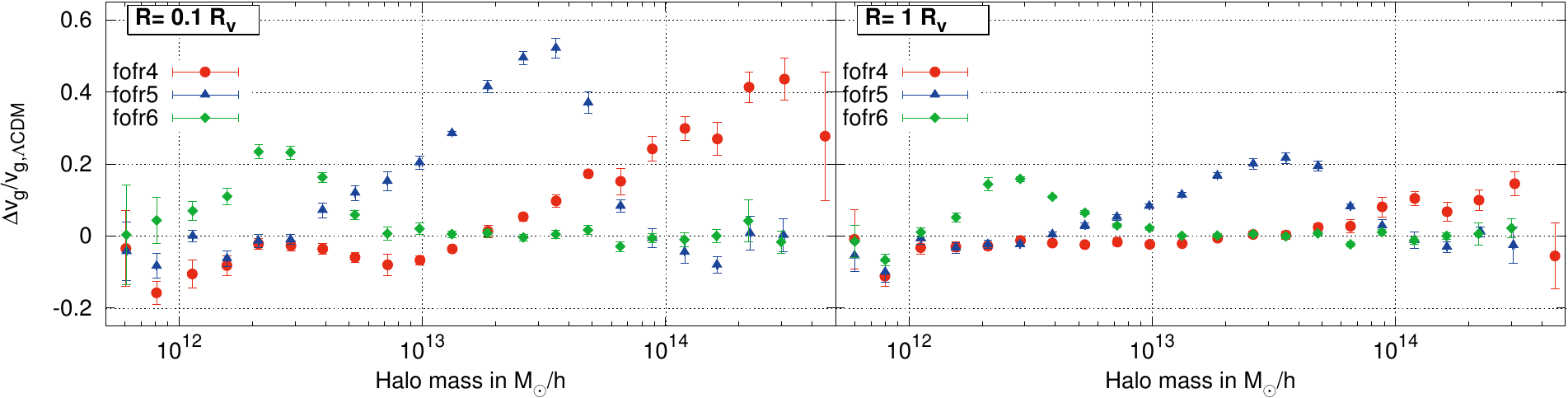}
  \\
  \includegraphics[width=.95\textwidth]{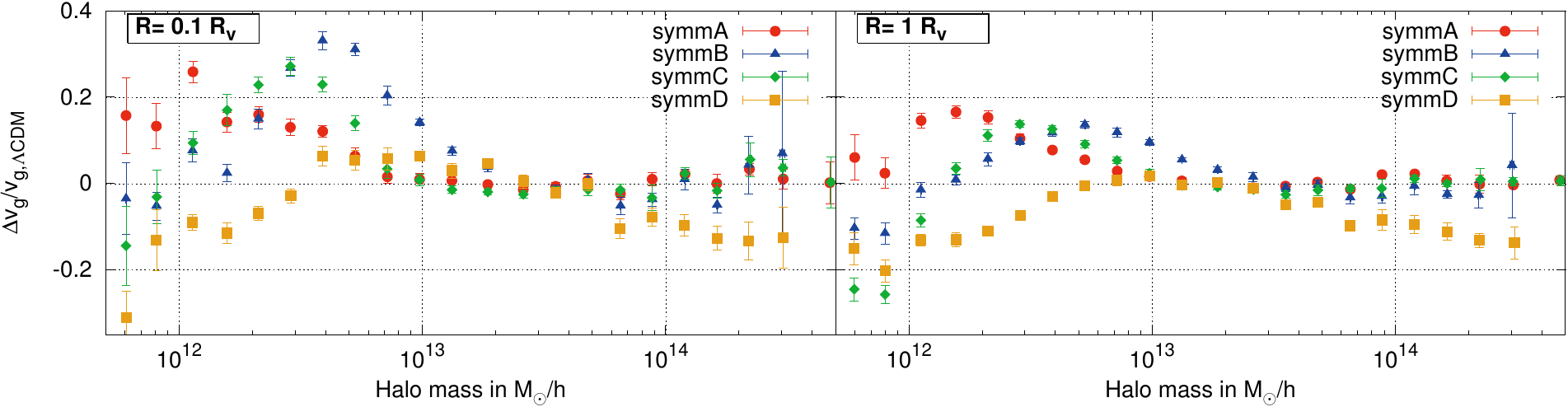}
  \caption{Relative deviation in gravitational redshift between the scalar-tensor models and $\Lambda$CDM as a function of halo mass for two different radial bins. The upper panels correspond to the $f(R)$ model and the lower to the symmetron.}
  \label{fig:grav_red_mass}
\end{figure*}

\subsection{Gravitational redshift}

The aim of the paper has been to study gravitational redshift.  As discussed in the introduction, in the family of alternative theories that we are treating in this paper, the scalar field does not affect the energy of photons, and thus, any difference in the gravitational redshift prediction will come from differences in the matter distribution.  Naively, one would expect that the presence of a fifth force will increase the clustering and, thus, produce \modi{deeper} potential wells, with the consequence of a\modi{n} increase in the gravitational redshift at all masses.  Our simulations show that the situation is a bit more complex.

Figure~\ref{fig:grav_red_radius} shows the relative deviation in the gravitational redshift $v_g$ with respect to $\Lambda$CDM as a function of radius.  The upper and lower panels show the $f(R)$ and symmetron results, respectively.  As previously in Fig.~\ref{fig:fifth_force_mass}, the vertical black line denotes and estimation of the resolution limit. Since the fifth force causes additional clustering, which affects the gravitational redshift profile of the clusters, we expect to find similar dependencies on the model parameters to those in the previous section. This clearly applies to the $f(R)$ results where we find, as expected, that the smaller $|f_{R0}|$, the smaller the deviation from $\Lambda$CDM. Especially in the $|f_{R0}|=10^{-6}$ case, the result is basically indistinguishable from $\Lambda$CDM. On the other hand, for the more extreme cases of $|f_{R0}|=10^{-4}$ and $|f_{R0}|=10^{-5}$, the relative deviation is much greater, in particular for small radii. In the case of the symmetron model, the imprint in the variation of the gravitational redshift profiles is not as strong as in the $f(R)$ results, which is a consequence of the fact that the symmetron is a more effective screening mechanism.  The maximum deviation in this case is around $15\%$ for the \textit{symm\_D} model for haloes with masses between $10^{14}$ and $10^{14.5}\,M_\sun h^{-1}$. As expected, this deviation decreases for later symmetry breaking times. In particular, the \textit{symm\_A} model with $z_{ssb}=1$ is basically indistinguishable from $\Lambda$CDM.  For most of the parameter sets and halo mass bins analysed, the results show a stronger gravitational redshift profile in the modified gravity models, following the expected behaviour.  However, the models \textit{symm\_B}, \textit{symm\_D} and \textit{fofr5} \modi{show} a signal that contradicts the naive expectations (i.e. show a negative correction towards a less important gravitational redshift).  Before discussing the reason for this happening, we briefly discuss the dependence of the gravitational redshift on mass.

The three halo mass bins displayed in the panels of Fig.~\ref{fig:grav_red_radius} already suggest that the amount of deviation in gravitational redshift depends on the cluster mass.  For instance, the \textit{fofr5} model shows the largest enhancement in gravitational redshift in the low mass bin, while the \textit{fofr4} becomes dominant for the highest mass haloes.  This mass dependency of the deviation in gravitational redshift is shown more clearly in Fig.~\ref{fig:grav_red_mass}, where we present the relative deviation with respect to Einstein gravity as a function of mass and for two different radii.  Here, we find mass ranges for which the deviation is greatest.  As previously noticed for the characteristic mass ranges noticed in Fig.~\ref{fig:fifth_force_mass}, the position of the peak depends on the model parameters.  For example, in the \textit{fofr6} data the maximum deviation is for haloes with masses around $2\times 10^{12}M_\sun/h$, and for \textit{fofr5} it is around $3\times 10^{13}M_\sun/h$.  The \textit{fofr5} model does not present a peak, but a continuous growth towards the highest mass haloes. Larger box sizes are needed to map the larger clusters and to confirm that the signal is also the same for this model.  Similarly, the symmetron curves also possess maxima whose positions correspond to the ones of the peaks in Fig.~\ref{fig:fifth_force_mass}. 

\section{Discussion}
\label{sec:discussion}

It is well known that scalar tensor models increase the clustering rate \citep[see for instance results from N-body simulations in][]{Brax2013a,Brax2012,Li2012}.  In particular, we find
\begin{itemize}
\item \textit{f(R) results:} The lower the value of $|f_{R0}|$, the smaller are the affected haloes, and the fifth force gets reduced. Since $f_{R0}$ controls directly, the deviation from $\Lambda$CDM in the action (see Eq.~(\ref{eq:fofr_eq3})) and the range of the field in vacuum is $\lambda_\phi= 1/\sqrt{V''(\phi)} \propto \sqrt{|f_{R0}|}$, which is the expected result.
\item \textit{symmetron results:} As expected from Eq.~\eqref{eq:geodesic_isis_symmetron2} a lower value of $a_{ssb}$ or a higher value of $\beta$ results in a stronger fifth force and a generally greater deviation from the $\Lambda$CDM cluster profiles. In this case the range of the field in vacuum $L$ was not altered, and consequently, the size of the affected haloes did not change dramatically. Still, we find lowering $a_{ssb}$ leads to bigger haloes being affected by the fifth force, since the symmetron force has more time to influence the matter clustering.
\end{itemize}

The most common expectation about the change in the gravitational redshift through this additional clustering is to observe an enhanced signal. This idea is motivated by the fact that additional clustering should lead to deeper potential wells, hence to a stronger gravitational redshift signal. Mostly, this is the major imprint observed in the $f(R)$ models studied, and the same behaviour was expected in the symmetron models. However, depending on the halo mass and model parameters, the matter density can be enhanced in the outskirts of the halo.  Keeping in mind the halo mass is effectively fixed in one mass bin, this result is not surprising as missing mass in one region has to be compensated by additional mass in another region. Such an altered halo density profile means the potential well becomes shallower in the central regions and a deeper in the outskirts. Consequently, the resulting deviation in gravitational redshift is negative. 
Still, this connection between clustering in the central regions and stronger gravitational redshift or clustering in the outer regions and a negative deviation does not explain the opposite imprint obtained for various parameter sets. Therefore, the question we tried to answer is what mechanism controls whether the clustering happens in the central region or in the outskirts of a dark matter halo?

To answer this question, the position of the additional clustering is analysed. This position is related to the radius of the maximal fifth force $R_{\text{max}}$ (see Fig.~\ref{fig:fifth_force_radius}) the following way. For $R\gtrsim R_{\text{max}}$, where the fifth force decreases because a lower matter density leads to a less affected $\phi$-value, the additional mass-flux towards the centre is greater than the supplementary inflow of mass towards $R_{\text{max}}$. For $R\lesssim R_{\text{max}}$, where the fifth force decreases due to a flattening of the halo density profile or because of screening, the opposite is the case. This means that the additional clustering happens around $R_{\text{max}}$. \modi{The} extent of the additional clustering depends on the overall strength of the fifth force and the steepness of the slopes around $R_{\text{max}}$. In summary, an $R_{\text{max}}$ close to the resolution limit leads to a deepening of the potential well and therefore to a positive deviation in gravitational redshift, whereas a greater value of $R_{\text{max}}$ indicates an additional clustering in the outskirts of the halo, hence a negative deviation with respect to $\Lambda$CDM.
Another way of phrasing it is that each set of model parameters results in a ``preferred density'', that is, a matter density for which the fifth force is maximal.

This reasoning can most easily be tested by observing the dependency of the fifth force and the deviation of the gravitational redshift for the \textit{symm\_D} model since $R_{\text{max}}\gtrsim 0.1 R_v$ in this case.
Here, clustering the centre is effectively weakened for clusters with mass $\gtrsim 10^{13}M_\sun h^{-1}$. Consequently, gravitational redshift is decreased in these haloes.
However, the explanation is supported by several additional features in the results. For instance, by decreasing $z_{ssb}$, the effect should still persist in a weaker form -- which it does in the \textit{symm\_B} data set.
A special position is taken by the \textit{fofr4} data: The increase in the number of massive clusters is immense, suggesting the same or even stronger merging of clusters. On the other hand, the force enhancement close to the centre has not yet reached its maximum, even for the largest haloes of the data sample (top left panel of Fig.~\ref{fig:fifth_force_mass}) opposed to the \textit{symm\_D} case, where this maximum lies around $10^{13}\,M_\sun h^{-1}$ for $R\approx 0.1\,R_v$. This means that, in the \textit{fofr4} case, the ``effective preferred density'' is even higher than the one present in the centre of the biggest clusters. Therefore, the clustering is dominant in the central parts, leading to a higher gravitational redshift as seen in Fig.~\ref{fig:grav_red_radius}. It is more appropriate to compare the \textit{symm\_D} with the \textit{fofr5} results owing to their greater similarity in the fifth force profiles. The fifth acceleration in the central region is, however, more present for haloes in the mass range $10^{13.5}-10^{14}M_\sun h^{-1}$ in the \textit{fofr5} data. For the next bigger haloes, $\vek{\ddot x_{\mathrm{Fifth}}}$ is negligible in the central regions in both models. This explains the resulting positive deviation in gravitational redshift for the first halo-mass bin mentioned and the negative deviation in the second (Fig.~\ref{fig:grav_red_radius}). Of course, the different other characteristics of the symmetron and the chameleon model (e.g. the time dependence of the fifth force) have to be taken into account if the data is to be analysed in more detail. The explanation provided can, after all, clarify the partially opposite results obtained.

\section{Conclusions}
\label{sec:conclusions}

We performed a set of eight high-resolution simulation runs with $512^3$ particles and a box length of $256\,\mathrm{Mpc\,h^{-1}}$ each. The data was used to analyse the density and gravitational redshift profiles of virialised clusters. For this purpose, the virialisation parameter was computed first for each halo with the inclusion of the scalar field energy.
Then, in order to calculate the gravitational redshift of all particles, the value of the gravitational potential at the particle's position was subtracted from the halo's minimal gravitational potential.
Ultimately, the modified gravity results could be compared to the $\Lambda$CDM values.
These steps were not only undertaken for the gravitational redshift and fifth force but also for the matter density, particle density, and velocity dispersion (not shown). These quantities allowed us to analyse and explain the gravitational redshift results. The overall finding from the numerical results is that in both analysed models, the deviation from $\Lambda$CDM can vary enormously depending on the choice of parameters and the analysed halo size.

The main goal of this work was to study gravitational redshift profiles of virialised haloes in both the symmetron model and chameleon $f(R)$-gravity using $N$-body simulations. To obtain the state of virialisation, several methods were analysed, and the most adequate one -- using the total acceleration of the particles -- was employed. This allowed us to calculate the virialisation parameter taking the energy of the scalar field into account. \modi{We note, however, that the qualitative results can be reproduced when including all haloes. This leads merely to an increase of noise, especially in the halo outskirts. In spite of this, we chose to only include the virialised haloes in our analysis to give a more conservative prediction.}

We found the results to be highly dependent on the halo mass, which means the consideration of multiple mass ranges is crucial when analysing observational data. In particular, three possible regimes were identified:
\begin{enumerate}[label=(\roman*)]
\item \textit{No deviation from $\Lambda$CDM.} If the halo is fully screened (either through self-screening or environmental screening) or is too small to affect the scalar field, no deviation in the density and, consequently, in the gravitational redshift profiles was observed.
\item \textit{Enhanced gravitational redshift.} This is the expected result since the fifth force leads to additional clustering in the centre of the halo. Additional matter in the central region  means a deeper potential well and, therefore, a positive deviation compared to $\Lambda$CDM.
\item \textit{Weaker gravitational redshift.} If the fifth force is screened in the central region of the halo but not in the outskirts, a negative deviation can appear. The additional force leads to a matter overdensity in the outer regions while -- with respect to the halo mass -- the inner regions are less dense than in the $\Lambda$CDM case.
\end{enumerate}
This shows that simply assuming the change in the gravitational redshift is equal to the maximum possible change in the gravitational constant, as frequently done when predictions are made, is not sufficient. Instead, the prediction obtained by $N$-body simulations as in this work should be used.

\citet{Croft2013} shows error predictions for the full SDSS/BOSS \citep{Aihara2011}, BigBOSS \citep{Schlegel2009}, and Euclid \citep{Laureijs2011} data sets. Accordingly, BigBOSS and Euclid should be able to map out the amplitude of the $v_g$ curve with $6.5\%$ and $4\%$ precision, respectively. These values were obtained using a different technique than our approach. Instead of taking the difference between the gravitational redshift of the central galaxy (BCG) and each particle in a cluster, the data was compared pairwise. As a result, they cannot be transferred directly. In addition, the binning of data in halo mass, which is necessary to detect signatures due to screened gravity, will naturally increase the error. Nevertheless, the planned new sky surveys will allow restriction of the parameter space of some modified gravity models.

\begin{acknowledgements}
M.B.G. would like to thank Michael Kopp for his useful comments on the manuscript. C.L. and D.F.M. thank the Research Council of Norway FRINAT grant 197251/V30.
D.F.M. is also partially supported by project CERN/FP/123618/2011 and CERN/FP/123615/2011.
The simulations were performed on the NOTUR Clusters \texttt{HEXAGON} and \texttt{STALLO}, the computing facilities at the University of Bergen and Troms\o~in Norway.
\end{acknowledgements}

\bibliographystyle{aa} 
\bibliography{references} 

\end{document}